\documentclass[aip,rsi,reprint]{revtex4-1} 
\usepackage{graphicx}
\usepackage{SIunits}
\usepackage[version=3]{mhchem}

\begin{document}


\title{Growing bubbles in a slightly supersaturated liquid solution} 



\author{Oscar R. Enr\'iquez}
\author{Christian Hummelink}
\author{Gert-Wim Bruggert}
\author{Detlef Lohse}
\author{Andrea Prosperetti}
\author{Devaraj van der Meer}
\author{Chao Sun}
\affiliation{Physics of Fluids Group, Faculty of Science and Technology, University of Twente, P.O. Box 217, 7500 AE Enschede, The Netherlands}


\date{\today}

\begin{abstract}
We have designed and constructed an experimental system to study gas bubble growth in slightly supersaturated liquids. This is achieved by working with carbon dioxide dissolved in water, pressurized at a maximum of $\unit{1}{\mega\pascal}$ and applying a small pressure drop from saturation conditions. Bubbles grow from hydrophobic cavities etched on silicon wafers, which allows us to control their number and position. Hence, the experiment can be used to investigate the interaction among bubbles growing in close proximity when the main mass transfer mechanism is diffusion and there is a limited availability of the dissolved species.
\end{abstract}


\maketitle 

\section{Introduction}

We study the growth of bubbles by gas diffusion in a liquid, which is the mass transfer mechanism when bubbles grow in a supersaturated solution. In experimental studies carried out so far\cite{Bisperink1994, Barker2002,Jones1999b}  the flow induced by the growing bubble on its surroundings might not be negligible. The consequence of this is larger growth rates than expected for pure diffusion. In a solution that is only very slightly supersaturated, bubbles should grow quasi-statically and hence exclusively by diffusion. In this paper, after briefly introducing the context of supersaturated liquids, we describe  an experimental system in which bubble growth can be studied under favourable conditions to isolate diffusion and where the number and position of bubbles can be controlled in order to study the interaction among them.

\subsection{Supersaturation and its occurrence}
The de-gassing of a supersaturated gas solution in a liquid takes place in a wide range of natural and industrial processes. Perhaps the most familiar examples are carbonated beverages, which have motivated a large amount of research on the physics and chemistry behind bubble formation, foaming and gushing in soda, beer and champagne\cite{Bisperink1994, Barker2002, LigerBelair2005,  Sahu2006, Lee2011}. Other examples include bubble growth in blood and tissues due to decompression sickness \cite{Chappell2006}, de-gassing of magmas during volcanic eruptions \cite{Sparks1978}, boiling-up of cryogenic solutions \cite{Kuni2002, Zhuvikina2002, Kuni2003}, production processes involving molten polymers, metals or glass \cite{Amon1984}, and ex-solution of gases during oil extraction \cite{Pooladi1999}.

As described by Henry's Law, the equilibrium (saturation) concentration, $c$, of gas in a liquid solution at a temperature $T$ is proportional to the partial pressure $P$ of the gas above the liquid:
\begin{equation}
\label{eq:HenrysLaw}
c=k_H P.
\end{equation}
Here $k_H$, the so-called Henry's constant, is specific to the gas-liquid pair and is a decreasing function of temperature. If the concentration $c_0$ of a gas-liquid solution, in thermodynamic equilibrium at a pressure $P_0$ and temperature $T_0$, is brought to a lower pressure $P_s$ and/or higher temperature $T_s$, it becomes supersaturated with respect to the equilibrium concentration $c_s=k_H(T_s)\,P_s$ at the new conditions. The excess amount of dissolved gas can be characterized in terms of the supersaturation ratio $\zeta$ defined by

\begin{equation}
\label{eq:supersatration}
\zeta = \frac{c_0}{c_s}-1.
\end{equation}
Clearly, supersaturation requires that $\zeta > 0$.  

Upon supersaturation, the excess gas must escape from the solution in order to re-establish equilibrium ($\zeta=0$). In a quiescent liquid this can be a rather slow process which involves diffusion through the free surface and formation of  gas bubbles that rise through the liquid and burst at the surface. A familiar example of this is the `going flat' of a carbonated drink that is left open, which can take a few hours. To further illustrate this example we can consider the case of Champagne wines, studied in depth by Liger-Belair\cite{LigerBelair2005}. In such drinks $\zeta\approx 5$ (with $c_s$ defined at $P_s=\unit{101}{\kilo\pascal}$). A $\unit{0.1}{\litre}$ glass of Champagne contains an excess of $\sim\unit{0.6}{\litre}$ of gaseous \ce{CO2} that, if left alone, will escape the liquid. Contrarily to what might be expected, it has been shown that only about 20$\%$ of the gas escapes inside the $\sim 2$ million bubbles of average diameter $\sim\unit{500}{\micro\meter}$ that will be formed. The other 80$\%$ leaves directly through the free surface \cite{LigerBelair2005}, although not without help from the mixing provided by the swarms of rising bubbles.

\subsection{Bubble nucleation}
The conditions necessary for gas bubbles to nucleate have been the object of substantial debate and study. Lubetkin\cite{Lubetkin2003} presented a list that illustrates the variety of arguments that have been put forward to explain the discrepancies between nucleation theory and experiments. The supersaturation ratio in the Champagne example is low compared to the theoretical predictions of $\zeta>1000$ in order for homogeneous nucleation to occur at room temperature \cite{Wilt1986}. Bubble growth below the homogeneous threshold requires the pre-existence of gas pockets\cite{Jones1999} (nucleation sites) with a radius equal to or larger than a critical value

\begin{equation}
\label{eq:crit_radius}
R_c=\frac{2\sigma}{P_s\zeta},
\end{equation}
with $\sigma$ the surface tension of the gas-liquid interface. This value is obtained by equating the concentration of gas in the liquid bulk (which immediately after supersaturation is equal to $c_0$) to the gas concentration at the surface of the gas pocket, given by $c_b=k_H(P_s+2\sigma/R)$. The second term in the parenthesis is the Laplace pressure jump due to a curved interface. A smaller gas pocket  will dissolve quickly since the concentration on its surface  exceeds $c_0$, causing an unfavourable concentration gradient. Larger ones, on the other hand, will induce a diffusive flow of the dissolved gas towards them and hence grow. In principle, nucleation sites might be provided by suspended particles, crevices in the container or free small bubbles. However, the latter are not stable. An undisturbed liquid which is left to rest will soon get rid of free bubbles either by dissolution or by growth and flotation\cite{Atchley1989}.
\begin{figure*}
 \includegraphics{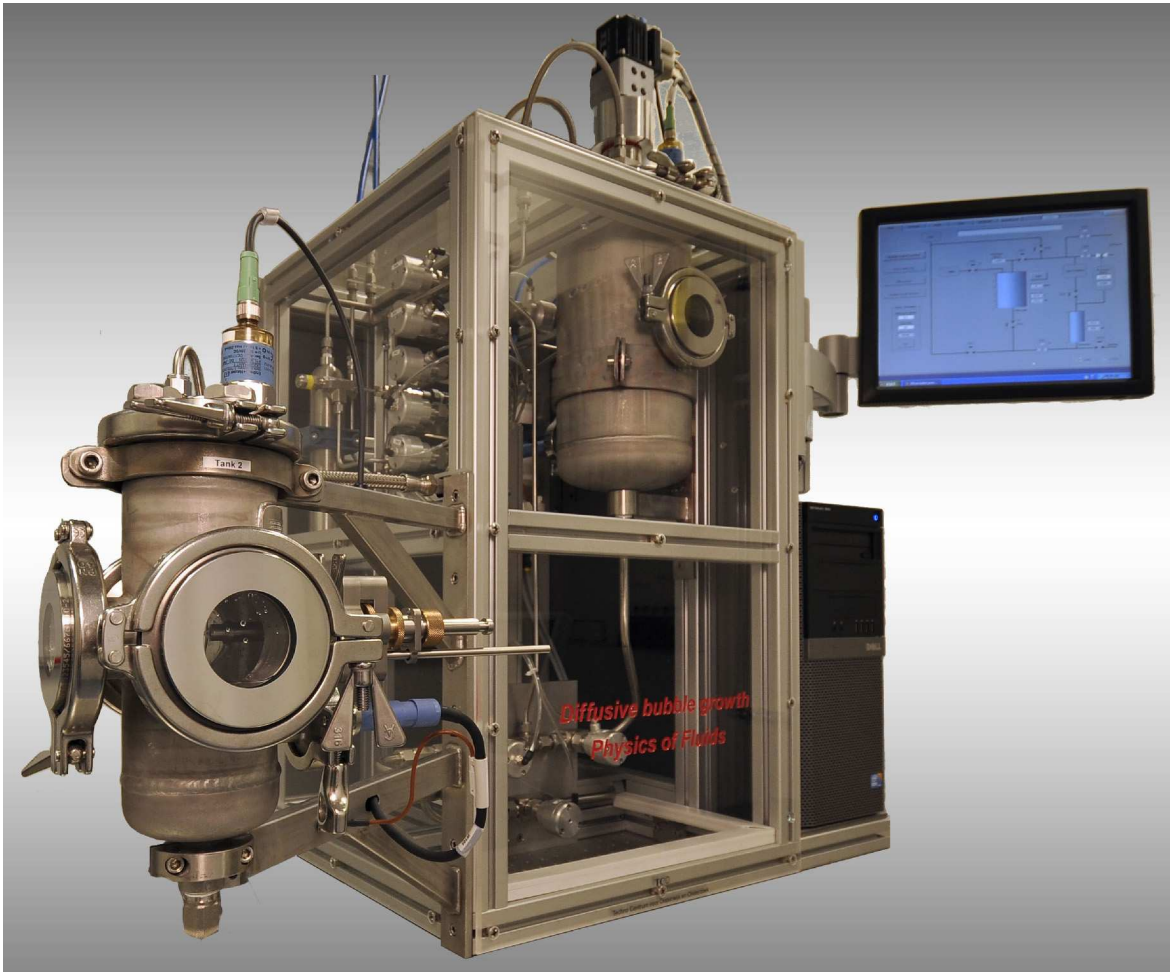}
 \caption{\label{fig1:setuppicture}(Color online) Photograph of the experimental system. The reservoir tank is located on the right-hand side and inside the frame. The observation tank is outside the frame in order to allow positioning of lights and cameras. The height of the frame is about 90 cm.}%
 \end{figure*}

 \subsection{Our experimental set-up}
It is our intention to study the growth of gas bubbles in a liquid with supersaturation $\zeta<1$, where bubble growth times are expected to be long. To our knowledge, there exist no previous experimental studies of diffusive bubble growth under such conditions. Previous studies have used values of $\zeta \sim 2$, which is comparable to the supersaturation of carbonated soft drinks\cite{Hey1994, Bisperink1994, Jones1999b}. We shall probe the limit of very slow degassing, first to observe the growth of a single bubble and then how bubbles interact when growing in mutual proximity while `competing' for a limited amount of available gas.

In this paper we describe an experimental system (fig.\ref{fig1:setuppicture}) designed to prepare a saturated solution and then supersaturate it by slightly decreasing its pressure (sections \ref{subsec:tanks} through \ref{subsec:control}). It is through accurate pressure control that we can achieve and maintain the small supersaturations desired for the experiments. Bubbles then grow in pre-determined positions provided by crevices in a specially prepared surface (sections \ref{subsec:substrates} and \ref{subsec:positioner}). This technique allows us to control the number of bubbles and the distance between them as we image their evolution digitally (section \ref{subsec:visual}). Finally, in sections \ref{sec:performance} and \ref{sec:outlook} we present the results of performance tests and the outlook of the experimental studies to be performed in the future.

\begin{figure*}
 \includegraphics{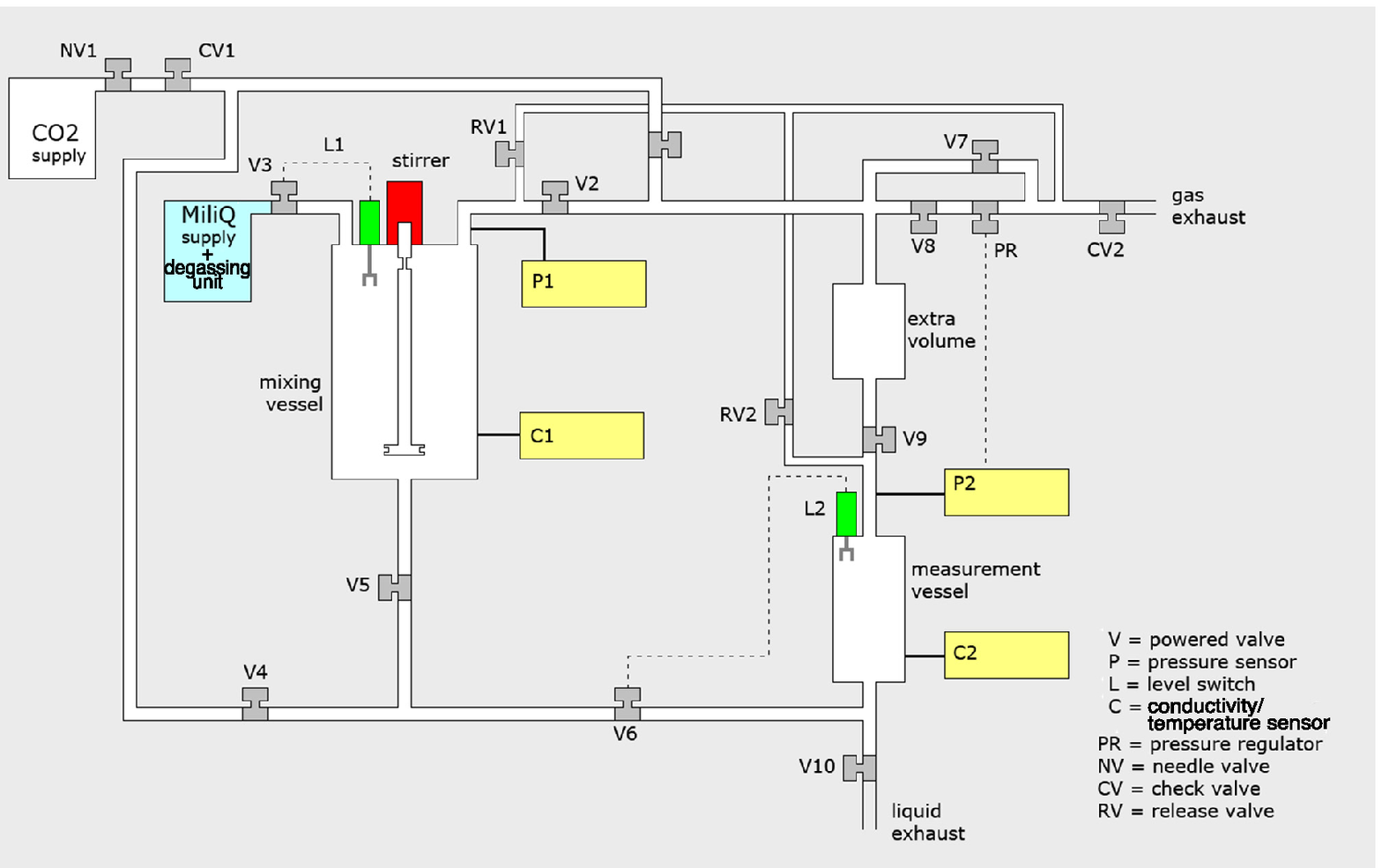}
 \caption{\label{fig2:setupdiagram}(Color online) Scheme of the setup indicating the location of valves, pressure controllers, and sensors. Here the position of the tanks is reversed with respect to the picture shown in figure 1.}%
 \end{figure*}
 
\section{Experimental set-up description}

\subsection{Stainless steel tanks}
\label{subsec:tanks}
 
The system (figures \ref{fig1:setuppicture} and \ref{fig2:setupdiagram}) is composed of two stainless steel tanks with volumes of 7 and 1.3 liters respectively. The larger one serves as a reservoir where a solution of  water saturated with gas can be prepared and stored. This mixture can be transferred to the smaller observation tank where the experiments in controlled bubble growth properly take place. A system of steel pipes and pneumatic valves connect the tanks to each other and to the water and gas sources as well as to the drainage system of the lab.

\begin{figure}
 \includegraphics{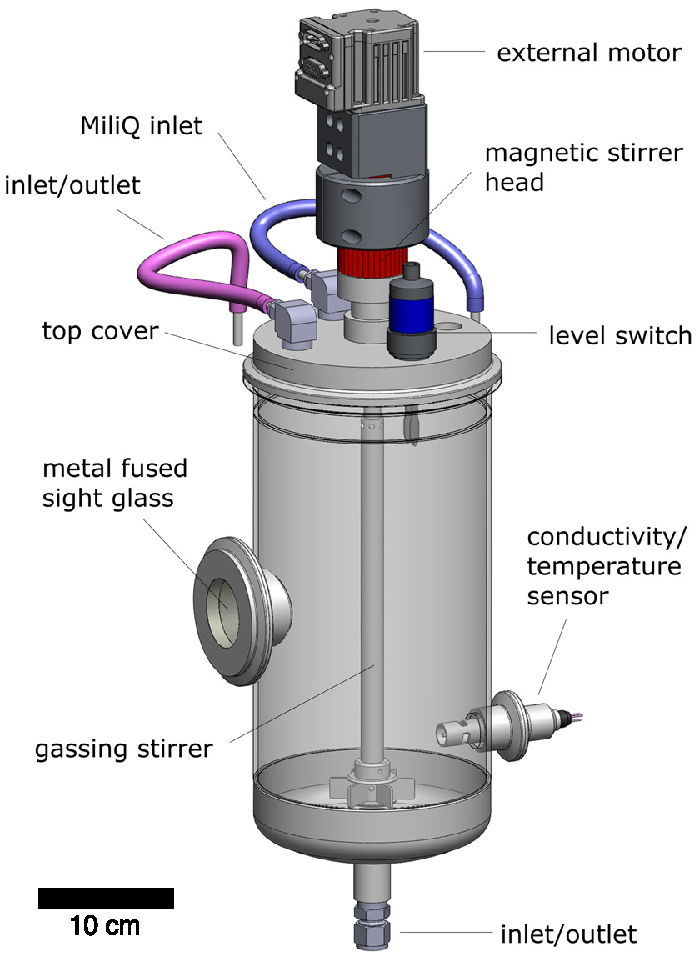}
 \caption{\label{fig3:mixtank}(Color online) The stainless steel reservoir tank used for preparing and storing a saturated mix of \ce{H2O} and \ce{CO2} at a maximum overpressure of 10 bar.}%
 \end{figure}

The tanks were manufactured from 3161 stainless steel (Het Noorden, Gorredijk, The Netherlands), and are certified for a working pressure of $\unit{1}{\mega\pascal}$. The reservoir (figure \ref{fig3:mixtank}) has a lateral flanged port for fitting a temperature/conductivity sensor (section \ref{subsec:concentration}), a lateral viewing window made of metal-fused  glass (Metaglas, Herberts Industrieglas), and a fluid inlet/outlet at the bottom. The plate that covers the top of the tank has fittings for a magnetic stirrer head (Macline mrk12, Premex Reactor AG), a level switch (Liquiphant FTL20, Endress+Hausser Inc.), a water inlet, and a gas inlet/outlet.

\begin{figure}
 \includegraphics{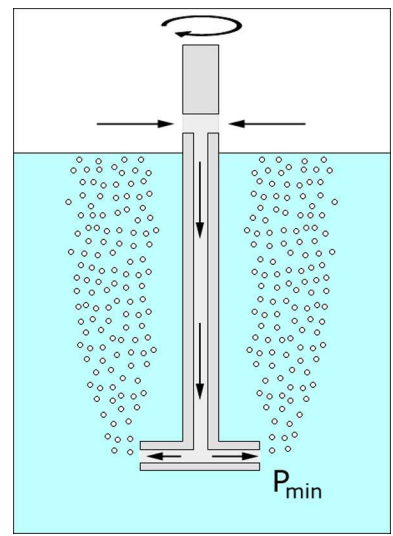}
 \caption{\label{fig4:mixer}(Color online) Sketch of the gassing mixer used. The rotation of the propeller blades creates low pressure zone. As a result, \ce{CO2} is sucked into the hollow stirrer axis and bubbled into the liquid through the end of the propeller blades}%
 \end{figure}

If we were to rely on natural diffusion for preparing the mixture of water saturated with gas, experimental waiting times would be extremely long. Hence, the reservoir tank is equipped with the aforementioned magnetic stirrer attached to a $\unit{285}{\milli\meter}$ gassing propeller (BR-3, Premex Reactor AG) and powered by an external motor (Smartmotor SM2315D, Animatics Corp). Figure \ref{fig4:mixer} shows how the mixer accelerates the saturation process. Rotation of the propeller blades creates a low pressure region around them. As a result, gas is sucked into the hollow stirrer axis and blown into the liquid through holes at the end of the propeller blades. With this system, the preparation of seven liters of saturated water takes less than one hour.

\begin{figure}
 \includegraphics{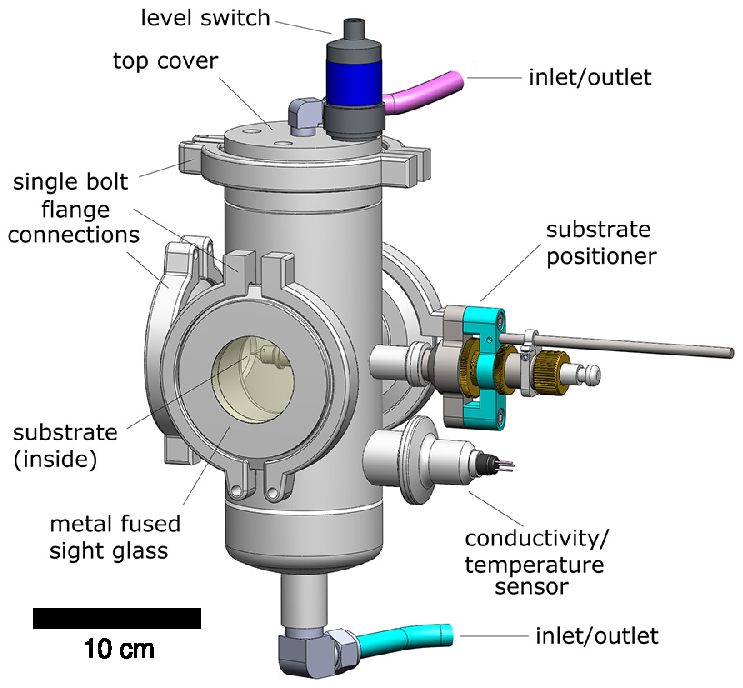}
 \caption{\label{fig5:visualtank}(Color online) The stainless steel observation tank. Part of the saturated mixture from the reservoir is transferred to this tank to be supersaturated by dropping the pressure in a controlled way. Bubbles grow on a sample held by the substrate positioner. The process is visualized through the windows.}%
 \end{figure}

The observation tank (figure \ref{fig5:visualtank}) has two lateral flanged ports: one for a temperature/conductivity sensor like the reservoir, and the other for introducing a specially designed tweezer (see section \ref{subsec:positioner}) designed to hold the substrates with nucleation sites for bubble growth. This tank has three viewing portholes also made of metal fused glass. These windows sit at 90\degree angles from each other and allow for illumination and visualization of experiments (see section \ref{subsec:visual}). The cover holds a level switch and a gas inlet/outlet. Water enters and exits through the bottom of the tank.

\subsection{Liquid and gas sources}
\label{subsec:water}   
Although in principle any transparent liquid-gas combination could be studied using this setup, the only configuration used up to now and in upcoming experiments is water with carbon dioxide. This mixture is convenient due to the high solubility of \ce{CO2} in water ($\sim 1.6\ \mbox{g}_{CO_2}/\mbox{kg}_{H_2O}$  at $T=20\degree\mbox{C}$ and $P=0.1$ \mega\pascal)  compared to other gases. 

We use ultra-pure water (MilliQ A10, Millipore) degassed in line by a vacuum pump (VP 86, VWR) coupled to a degassing filter (Millipak 100). The \ce{CO2} is provided by Linde Gas with 99.99\% purity.

\subsection{Pressure control}
\label{subsec:pressure}
As stated in Henry's Law, the quantity of a gas that can be dissolved into a liquid is directly proportional to the partial pressure of the gas above the liquid. The proportionality constant (Henry's constant) reflects the solubility of the gas-liquid pair and is a function of temperature. Therefore, by altering either pressure or temperature of a saturated solution it is possible to take it to an under or supersaturated state. In our experiments, we control supersaturation by dropping the pressure in the observation tank and keeping the temperature constant. 

The pressure at which the liquid is saturated in the mixing tank is controlled through a regulator on the \ce{CO2} line of the laboratory which has a maximum working pressure of 1 \mega\pascal. The value inside the tank is measured with a pressure transmitter (Midas C08, Jumo GmbH) which is read out by a multiparameter transmitter (ecoTrans Lf03, Jumo GmbH) that communicates with the general control interface (see section \ref{subsec:control})

\begin{figure}
 \includegraphics[width=6cm, angle=270]{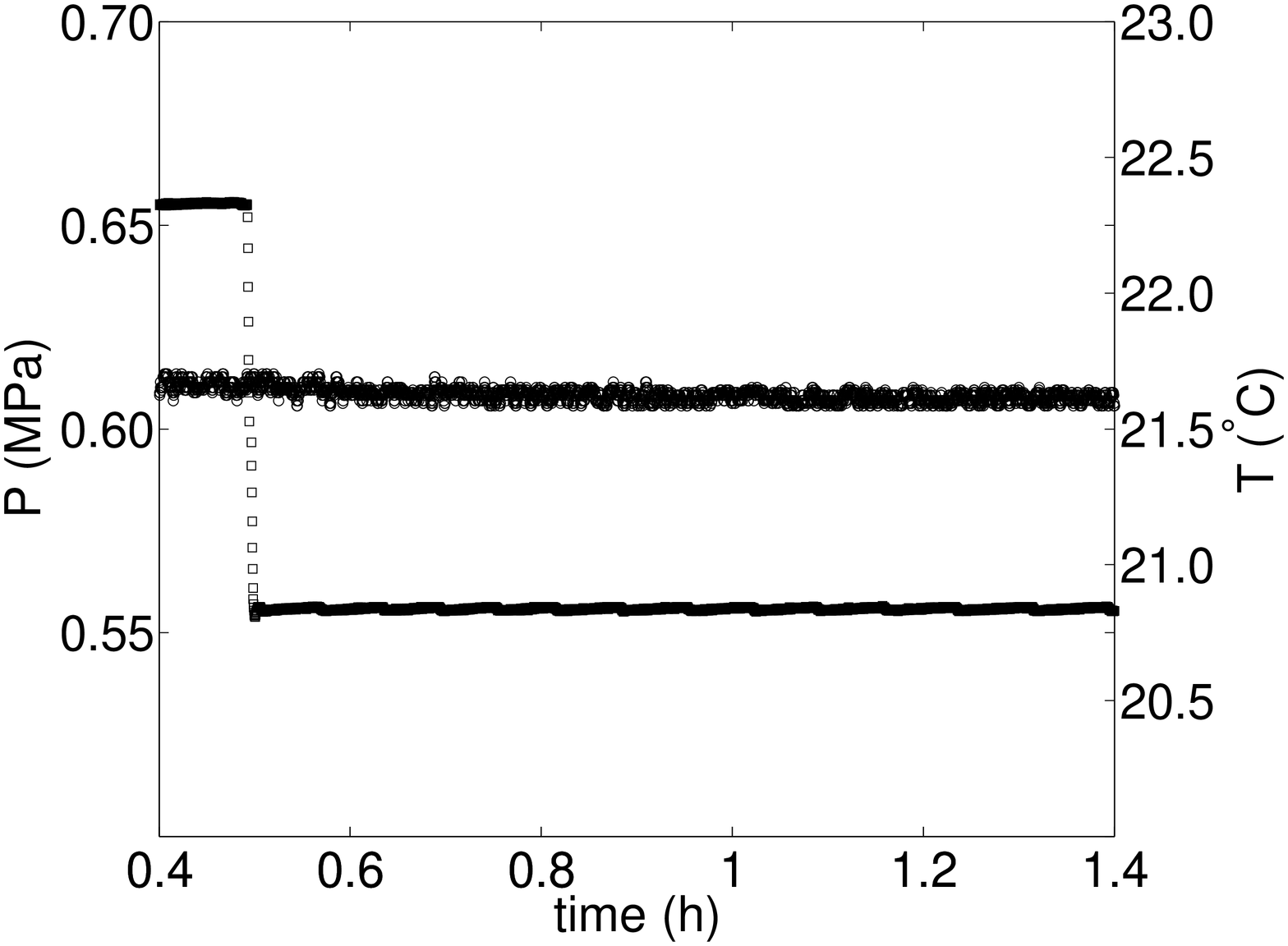}\\
 
 \caption{\label{fig:pressure} Example of a time series of pressure (squares) and temperature (circles) measurements in the observation tank during an experiment. The pressure was decreased by $\unit{0.1}{\mega\pascal}$ from the saturation condition, and kept at a constant value afterwards.}%
 \end{figure}

The pressure in the observation tank is measured and controlled by a pressure regulator (P-502C, Bronkhorst) and flow controller (F-001AI, Bronkhorst). The pressure regulator has a pressure range of 0.02-1 \mega\pascal\ with a measurement error of 5 \kilo\pascal. The flow controller has a working pressure range of 0.1-1 \mega\pascal\ and its flow range is 10-500 \milli\litre\per\minute. Since this type of control is based on a certain controlled volume, an extra volume of 500 \milli\litre\ is placed between the measurement vessel and the flow controller to permit a smooth regulation of the pressure. Figure \ref{fig:pressure} shows the pressure in the observation tank during a bubble growth experiment (see section \ref{subsec:bubble_epx}) where the pressure is dropped from an initial saturation state and kept constant as the solution degasses.

\subsection{Monitoring concentration}
\label{subsec:concentration}
Carbon dioxide reacts with water to form carbonic acid (H$_2$CO$_3$) which is unstable and dissociates into roughly equal amounts of hydrogen (H$^+$) and bicarbonate (HCO$^-_3$) ions. The amount of each chemical species and their molar conductivity will determine the general conductivity of the solution\cite{Light1995}. This property is used to monitor the concentration of \ce{CO2} during the saturation process in the mixing tank. For preparing the solution, the water filled tank is pressurized with \ce{CO2} and the mixer turned on. The rise in conductivity is immediately detected by the sensor and it saturates after some time. Measuring this property, therefore, serves as an indicator that saturation has been reached. It is assumed that after 10 minutes of measuring a stable conductivity value the desired state is achieved (see section \ref{subsec:solution_exp}).

In the case of the observation vessel, the measurement of conductivity serves as a qualitative indicator of the amount of \ce{CO2} present in the mix. Upon de-pressurization gas diffuses out of the solution. In the absence of significant mixing -as is the case during experiments- the main mechanism of gas exsolution is diffusion out of the free surface. Therefore a concentration gradient is established through the mixture and the conductivity measurement close to the bottom of the tank is no longer representative of the overall concentration of \ce{CO2}.

Conductivity and temperature of the liquid in both vessels are measured with  2-electrode conductivity sensors with integrated Pt100 temperature probes  (Condumax CLS16, Endress+Hausser Inc.) located near the bottom of each tank (fig.\ref{fig2:setupdiagram}). Knowing the temperature during experiments is necessary in order to correctly quantify the amount of supersaturation by knowing the correct value of Henry's constant. To avoid significant temperature variations, a hose (not shown in fig.\ref{fig1:setuppicture}) is wrapped around both thanks, through which water circulates at a temperature controlled with a refrigerated/heated circulator (Julabo, F25HL). Figure \ref{fig:pressure} shows the temperature in the observation tank during an experiment.

\subsection{Control and user interface}
\label{subsec:control}
The elements of the experiment that require electronic control are the magnetic stirrer, the level switches, the pneumatic powered valves and the flow regulator used for gradual pressure release. Control of these elements, together with data acquisition of the sensors, is done through a combination of programmable logic controllers (PLCs) (BC9120, Bus Terminal Controller, Beckhoff)  and a graphical user interface built in National Instruments LABVIEW which communicates with the PLCs and the data transmitters from the sensors.

\subsection{Subtrates for bubble nucleation}
\label{subsec:substrates}
Controlling the positions where bubbles grow is of paramount importance for studying the interaction of bubbles growing near each other as they `compete' for the gas available in dissolution. For this purpose we use silicon wafers of area around $\unit{1}{\centi\square\meter}$ with micron sized pits (of radius $R_{pit}=10-\unit{50}{\micro\meter}$ and depth $\sim\unit{30}{\micro\meter}$) that function as nucleation sites. The substrates are fabricated in a clean room using soft lithograpy and Reactive-Ion-Etching (RIE) techniques which allow to create pits with a minimum diameter of a couple microns and depths of a few tens of microns. In order to ensure that gas will be entrapped inside the pits after being submerged in water, the last step in the micro-manufacturing process is to create a super-hydrophobic `black silicon'\cite{Jansen1995} structure in the bottom of the pits. This guarantees that the air pockets in the cavities will be stable and henceforth work as nucleation sites for bubbles to grow upon de-pressurization. The feasibility and stability of such hydrophobic cavities as nucleation sites has been successfully tested by Borkent \textit{et al.}\cite{Borkent2009}.

\subsection{Substrate holder}
\label{subsec:positioner}
\begin{figure*}
 \includegraphics{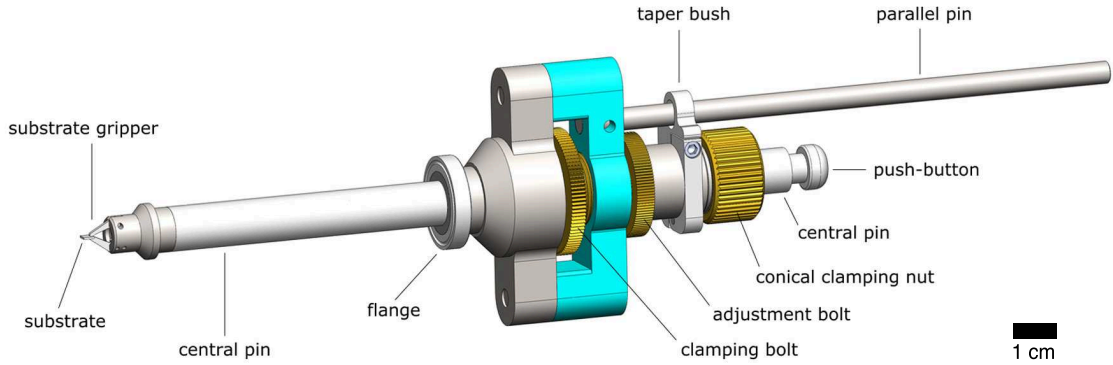}
 \caption{\label{fig7:positioner}(Color online) The substrate holder. The section to the left of the flange is introduced in the observation tank. The bolts on the right hand side are used to adjust the position of the substrate from outside the tank and keep it fixed firmly against the high pressure inside. The substrates can be held horizontally, vertically or at any angle in between by rotating the central pin.}%
 \end{figure*}

The substrates are introduced and held in the observation tank using the holder shown in figure \ref{fig7:positioner}. This device keeps the substrate at a level where it is visible through the windows and separated ($\unit{\sim5}{\centi\meter}$ from the walls, in order to avoid interaction with bubbles that might grow there.  It consists of a set of tweezers (`substrate gripper' in the figure) with one fixed and one mobile lever. The mobile one is actuated via the push button on the right-hand side and a spring mechanism that runs inside the central pin and keeps it in a closed position by default. The central pin can be slid back and forth and rotated by hand by loosening the conical clamping nut which will keep it fixed against the pressure in the tank. The clamping and adjustment bolts allow for a fine positioning along the direction of the parallel pin on which the guiding taper bush is mounted. 

The substrate is mounted on the tweezers outside the tank and then introduced through a flanged port (see figure \ref{fig5:visualtank}) and secured with a single-bolt clamp. In this and all other flange connections,  o-rings are used to ensure the water and air tightness of the system.

\subsection{Visualization}
\label{subsec:visual}
Images are taken using a  long distance microscope objective (K2/SC, Infinity) with a maximum working distance of 172 \milli\meter\ and a CCD camera (Flowmaster, LaVision) with a resolution of 1376 x 1040 pixels. When experiments are done with the substrate in a horizontal position diffuse backlighting is used. If the substrate is held vertically light is reflected onto it though a half mirror in front of the microscope.

\section{First experiments}
\label{sec:performance}

\subsection{Preparing a saturated solution}
\label{subsec:solution_exp}
Firstly, we have tested how effective our system is for preparing a saturated water-\ce{CO2} mixture and to what extent the measurement of conductivity serves as an indicator of the concentration. As mentioned in section \ref{subsec:concentration}, \ce{CO2} in water dissociates according to
\begin{equation}
\ce{CO2_{(aq)} + H2O <-> H^+ +HCO3^-},
\end{equation}
with overall dissociation constant $\ce{K1}=4.22\times 10^{-7}$ at $\unit{21}{\degree}$C. The molar conductivities (\ce{\Lambda^o_i} ) of the hydrogen and bicarbonate ions and their concentrations will determine the conductivity of the solution. The contributions of the dissociation of \ce{H2O} and \ce{HCO3^-} can be safely neglected. Hence, the conductivity ($S$) can be calculated as:
\begin{equation}
\label{eq:conductivity}
S=(\ce{\Lambda^o_{\ce{H^+}}} +\ce{\Lambda^o_{\ce{HCO3^-}}}) \ce{K1}[\ce{CO2_{aq}}],
\end{equation}
where the concentration of carbon dioxide (\ce{[CO2_{aq}]}) is expressed in $\unit{}{\mole/\cubic\meter}$, $\ce{\Lambda^o_{\ce{H^+}}}=\unit{348.22\times 10^{-4}}{\siemens\ \square\meter}$ and $\ce{\Lambda^o_{\ce{HCO3^-}}}=\unit{40.72\times 10^{-4}}{\siemens\ \square\meter/\mole}$ at $\unit{21}{\degree}$C. Such a way of determining the conductivity is expected to work only for low concentrations of dissolved \ce{CO2}, since due to its weak acidity, the ion concentration is not necessarily linear with \ce{[CO2_{aq}]}\cite{Light1995}.

\begin{figure}
 \includegraphics[width=7.5cm]{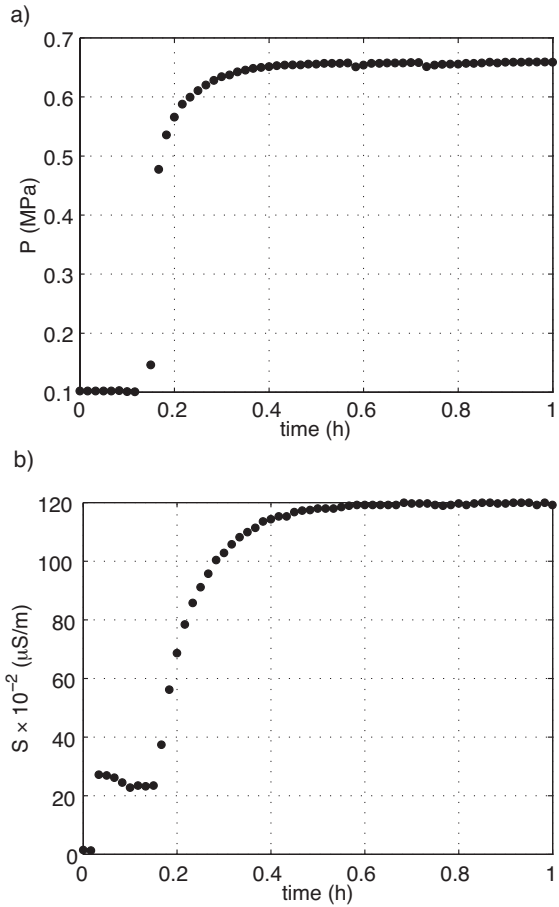}
 \caption{\label{fig:mix} Pressure (a) and conductivity (b) in the reservoir tank during the filling, pressurizing and mixing of the solution until saturation. The first jump of the conductivity corresponds to the immersion of the sensor in water as the reservoir is filled. Once full, the pressurization and mixing start. The solution reaches saturation after about half an hour.}%
 \end{figure}

\begin{figure}
 \includegraphics[width=6cm, angle=270]{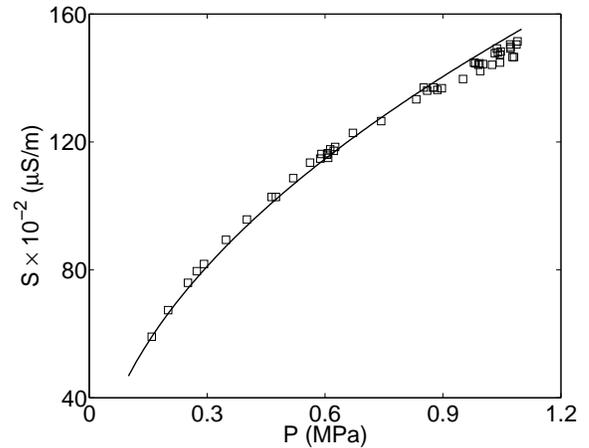}
 \caption{\label{fig8:conductivity} Conductivity of the saturated $\ce{H2O} + \ce{CO2}$ solution. Squares are the values measured by the conductivity sensor after the preparation procedure. The solid line is obtained with equation \ref{eq:conductivity}. }%
 \end{figure}

Tests were performed by filling the reservoir tank with water, leaving about 3 \centi\meter\ of head space for gas. Subsequently, and after the  pressure regulator of the gas line had been set to the desired pressurization value, the \ce{CO2} inlet valve was opened. At this time we also started the mixing propeller with a speed of 900 rpm so that \ce{CO2} is forced into the liquid through the mechanism described in figure \ref{fig4:mixer}. The gas inlet valve is kept open throughout this procedure in order to keep the pressure rising as gas dissolves into the liquid. We monitored the conductivity measurement of the sensor as it rose throughout the process. When its value did not change for 10 minutes we considered the solution to be stable, which was achieved after around 30 minutes of mixing as shown in figure \ref{fig:mix}. Henry's constant was computed using the van 't Hoff equation for its temperature dependence:\cite{Atkins1978}
\begin{equation}
k_H(T)=k_H(\Theta)\mbox{exp}\left[C\left(\frac 1T-\frac{1}{\Theta}\right)\right],
\end{equation} 
where $\Theta$ is the standard temperature ($\unit{298}{\kelvin}$) and $C=\unit{2400}{\kelvin}$ for the case of \ce{CO2}\cite{Lide1995}. The temperature during experiments, $T$ was $\sim\unit{21}{\degree}$C, giving $k_H=\unit{3.79\times 10^{-7}}{\mole/\kilo\gram\cdot\pascal}$. We then use Henry's law to calculate \ce{[CO2_{aq}]} at the experimental pressure, and introduce this value into equation \ref{eq:conductivity} to calculate the conductivity. Figure \ref{fig8:conductivity} shows the measured and the calculated values for saturation (absolute) pressures going from 2 to 11 bar. Agreement is very good until around 9 bar, when presumably the concentration of \ce{CO2} can no longer be considered as `low'.

\subsection{Growing bubbles}
\label{subsec:bubble_epx}

After making sure that our method to prepare the saturated solution is effective, we tested the bubble growing process from a single cavity and a pair of cavities. The typical procedure of an experiment is the following: the whole system is flushed with \ce{CO2} in order to expel atmospheric gasses. A saturated solution is prepared in the reservoir tank and part of it is transferred to the observation tank where the substrate with artificial nucleation sites was previously mounted on its holder. The filling is done by first pressurizing the tank to the same level as the reservoir in order to avoid a sudden, high pressure inflow of (supersaturated) water. The valves that connect the bottom of the two tanks (V5 and V6 in fig.\ref{fig2:setupdiagram}) are then open. Water flows slowly into the observation tank driven by the slightly higher positioning of the reservoir. The level switch (L2) closes the valves, thus ensuring that the tank is always filled to the same level. After this procedure we wait for half an hour to let water become completely stagnant. Then the experiment can start.  

The mix in the observation tank is supersaturated by reducing the pressure of the gas above it. Since we want to study diffusive growth without effects like inertia or streaming which appear when bubbles grow quickly in succession as in the case of, \textit{e.g.}, Champagne\cite{LigerBelair2005}, the pressure is dropped only 5 to 20\% from the absolute saturation pressure, giving a corresponding range of supersaturation $\zeta=0.05-0.25$. The critical radius (eq. \ref{eq:crit_radius}) for a gas pocket to grow under the smallest $\zeta$ considered is $\sim\unit{5}{\micro\meter}$, which means that hydrophobic pits of radius $10-\unit{50}{\micro\meter}$ are very well suited as nucleation sites under our experimental conditions. 

\begin{figure}
 \includegraphics{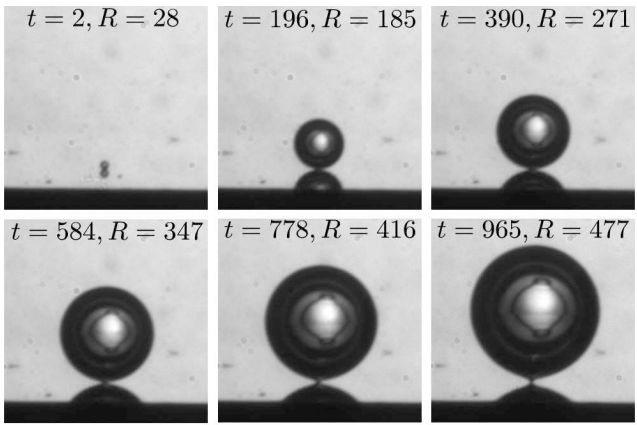}
 \caption{\label{fig9:bubble_images} Bubble growing on a substrate positioned horizontally with a single pit of radius $\unit{10}{\micro\meter}$ after a pressure drop from 0.65 to $\unit{0.6}{\mega\pascal}$ corresponding to a supersaturation $\zeta=0.08$. Time is expressed in seconds and radius in micro meters.}%
 \end{figure}

\begin{figure}
 \includegraphics[width=6cm, angle=270]{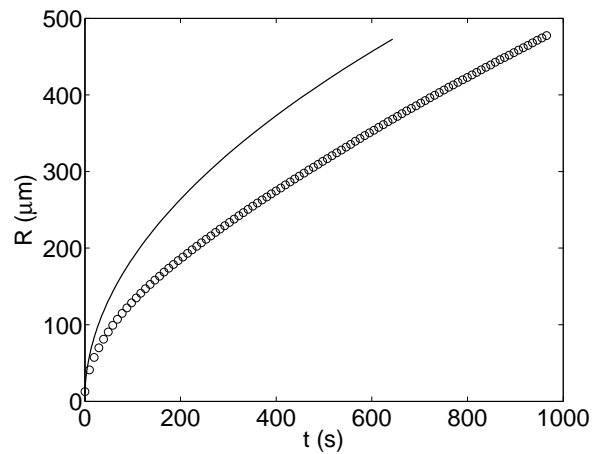}
 \caption{\label{fig10:radius_vs_time} Radius evolution of the bubble shown in figure \ref{fig9:bubble_images} $\circ$ experimental measurements, --- theoretical solution for a bubble growing by gas diffusion in an unbounded medium. }%
 \end{figure}

Figure \ref{fig9:bubble_images} shows a bubble growing from a pit with a $\unit{10}{\micro\meter}$ radius after a pressure drop from 6.5 to 6 bar and figure \ref{fig10:radius_vs_time} shows its radius as a function of time. The size at which the bubble detaches is determined by a competition between buoyancy pulling upwards and surface tension, pulling downwards. It is known as the Fritz radius\cite{Fritz1935} and for a spherical bubble is given by:
\begin{equation}
\label{eq:Fritzradius}
R_{Fritz}=\left(\frac 32 \frac{\sigma R_{pit}}{\rho g}\right)^{1/3}.
\end{equation}
The last observed radius of the bubble in figure \ref{fig9:bubble_images} before detachment was $\sim\unit{477}{\micro\meter}$ which is $\sim\unit{5}{\micro\meter}$ larger than the Fritz value for such a pit (with $\sigma=\unit{0.069}{\newton\per\meter}$ due to the presence of \ce{CO2}). The discrepancy is about $1\%$ and could be due to the fact that the tracking method assumes a spherical bubble, and at this point the latter is slightly deformed, or to small deviations in the pit's radius.  However, regardless of the cause, this is the maximum error incurred in the image processing, which we consider acceptable.

Two interesting things can already be pointed out from figure \ref{fig9:bubble_images}. The first is the fact that the bubble took more than 15 minutes to grow to a radius of $\unit{\sim0.5}{\milli\meter}$, which makes it a safe assumption to say that the only mass transfer mechanism present was diffusion. The second is that its growth was much slower than the solution of Epstein and Plesset\cite{Epstein1950} for a bubble growing under such supersaturation, which, as expected from diffusive processes predicts a $R\sim\sqrt{t}$ evolution. Their solution assumes an unbounded bubble in an infinite medium, so the slowing down is probably due to the presence of the substrate where the nucleation site is located. This feature will be studied systematically with the present apparatus.

After the bubble detaches, another one starts growing from the same place. As far as we have observed, this sequence continues for at least 12 hours. The amount of undesired bubbles growing on the walls of the tank is small, which means that the water is mainly degassing by diffusion through the gas-water interface above. The stratification provoked by the escape of \ce{CO2} from the surface is gravitationally stable and therefore will not give rise to density-driven convection unlike the opposite case of an undersaturated liquid pressurized with gas from above\cite{Lindeberg1997}. This considered, with the nucleation site sitting $\approx\unit{15}{\centi\meter}$ below the surface, and the diffusivity of \ce{CO2} in water being $D=\unit{1.97\times10^{-9}}{\square\meter\per\second}$, the time for the diffusive penetration length ($\delta_d=2\sqrt{Dt}$) to reach the bubble site should be about 1 month. In practice, the local concentration around the nucleation site will eventually drop to a level where the radius of the pit is less than $R_{crit}$ and the site will become inactive. However, we expect that the bubble growth sequence can continue for a couple of days after the initial pressure drop.

\begin{figure}
 \includegraphics{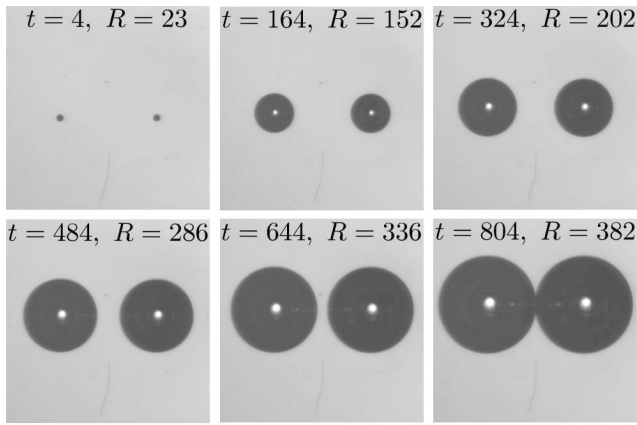}
 \caption{\label{fig:duo_bubble} Two bubbles growing on a substrate positioned vertically with two pits of radius $\unit{10}{\micro\meter}$, separated $\unit{760}{\micro\meter}$ after a pressure drop from 0.65 to $\unit{0.6}{\mega\pascal}$. Time is expressed in seconds and radius (values correspond to left-hand bubble) in micro meters.}%
 \end{figure}

Finally, we have tested the case of two bubbles growing close to each other. Figure \ref{fig:duo_bubble} shows two nucleation sites, separated $\unit{760}{\micro\meter}$ from which bubbles grow after an equal pressure drop to figure \ref{fig9:bubble_images}. In this case the substrate was positioned vertically and lit from the front through a half mirror. The growth of the pair of bubbles is slightly slower than the single bubble case, suggesting that each one of them influences the growth rate of the other. 

\section{Summary and outlook}
\label{sec:outlook}
We have developed an experimental system with which bubble growth by gas diffusion can be studied quantitatively. The method used to prepare a saturated solution of \ce{CO2} in water by pressurizing and mixing in a reservoir tank while monitoring the electrical conductivity has been shown to be effective. The position of bubbles growing when the solution is supersaturated by dropping its pressure can be accurately controlled using hydrophobic pits on silicon wafers. First experiments with a single bubble and a pair of them suggest that diffusion is indeed the only mass transfer mechanism in action.

The next step is to take a close look at the sequential growth of bubbles from a single nucleation site in order to understand the differences with the growth of an unbounded bubble. Afterwards we will investigate how multiple bubbles interact when growing in close proximity under low supersaturation conditions.

\begin{acknowledgments}
The authors would like to acknowledge the technical center (TCO) of the Faculty of Science and Technology from the University of Twente for their invaluable help in building the experiment and implementing the control. Many thanks to David Fern\'andez-Rivas, Stefan Schlautmann, and Meint de Boer with their assistance in the micro-pit fabrication. This work is part of the research programme of the Foundation for Fundamental Research on Matter (FOM), which is part of the Netherlands Organisation for Scientific Research (NWO). We also acknowledge financial support from the National Council for Science and Technology (CONACYT, Mexico), and Shell.
\end{acknowledgments}


%

\end{document}